\documentclass[twocolumn,noshowpacs,pre,preprintnumbers,superscriptaddress,amsmath,amssymb,floatfix]{revtex4}
\usepackage[caption=false]{subfig}
\usepackage{graphicx}
\usepackage{color}
\usepackage{placeins}
\usepackage{tikz}
\usepackage{mathtools}
\usepackage{amsthm}
\usepackage{romannum}

\newcommand{\er}{Erd\H{o}s-R\'{e}nyi}

\newcommand{\fij}{f_{i,j}}

\newcommand{\ktot}{\langle k_{total} \rangle}

\newcommand{\kinter}{\langle k_{inter} \rangle}

\newcommand{\mkkij}{\langle k_{i,j} \rangle}

\begin{document}

\title{Spreading of localized attacks on spatial multiplex networks with a community structure}
\author{Dana Vaknin}
\affiliation{Bar-Ilan University, Ramat Gan, Israel}
\author{Bnaya Gross} 
\affiliation{Bar-Ilan University, Ramat Gan, Israel}
\author{Sergey V.Buldyrev}
\affiliation{Yeshiva University, New York, USA}
\author{Shlomo Havlin}
\affiliation{Bar-Ilan University, Ramat Gan, Israel}
\date{\today}

\begin{abstract}

We study the effect of localized attacks on a multiplex spatial network, where each layer is a network of communities.
The system is considered functional when the nodes belong to the giant component in all the multiplex layers.
The communities are of linear size $\zeta$, such that within them any pair of nodes are linked with same probability, and additionally nodes in nearby communities are linked with a different (typically smaller) probability. This model can represent an interdependent infrastructure system of cities where within the city there are many links while between cities there are fewer links.
We develop an analytical method, similar to the finite element method applied to a network with communities, and verify our analytical results by simulations.
We find, both by simulation and theory, that for different parameters of connectivity and spatiality --- there is a critical localized size of damage above which it will spread and the entire system will collapse.

\end{abstract}
\maketitle


In recent years, due to the advances in technology, many systems have become more and more integrated and interdependent. 
This dependence between these systems can cause a spread of damages, and lead to a cascade of failures and even entire system collapse.
Therefore, many studies have been carried out to analyze cascading failures in interdependent networks \cite{buldyrev-nature2010,parshani-prl2010,gao-pre2012,baxter-prl2012,dedomenico-prx2013,bianconi-pre2013,son-epl2012}.
Further, in many real systems such as power grids and transportation systems, the links are of typical relatively short length due to the embedding in space. 
In such spatial systems, the initial failures or attacks can be localized to a specific region.
Recent studies show that in different cases of spatial interdependent networks, localized attacks are significantly more damaging than random attacks \cite{wei-prl2012,berezin-scireports2015,shao-njp2015,vaknin-njp2017}.
In addition, many real networks have a modular structure, such as biological networks \cite{bullmore-naturereviews2012} and many infrastructure systems \cite{eriksen2003modularity,guimer-nas2005}. Therefore, recent studies have explored and compared the robustness of individual and interdependent modular non-spatial systems \cite{bagrow-networkscience2015,shekhtman-njp2015,dong-pnas2018,gross2019interconnections}.	
Our study combines for the first time, three ubiquitous features of real complex systems --- interdependence, spatiality and modularity.

\begin{figure}
	\centering
	\includegraphics[width=0.65\linewidth]{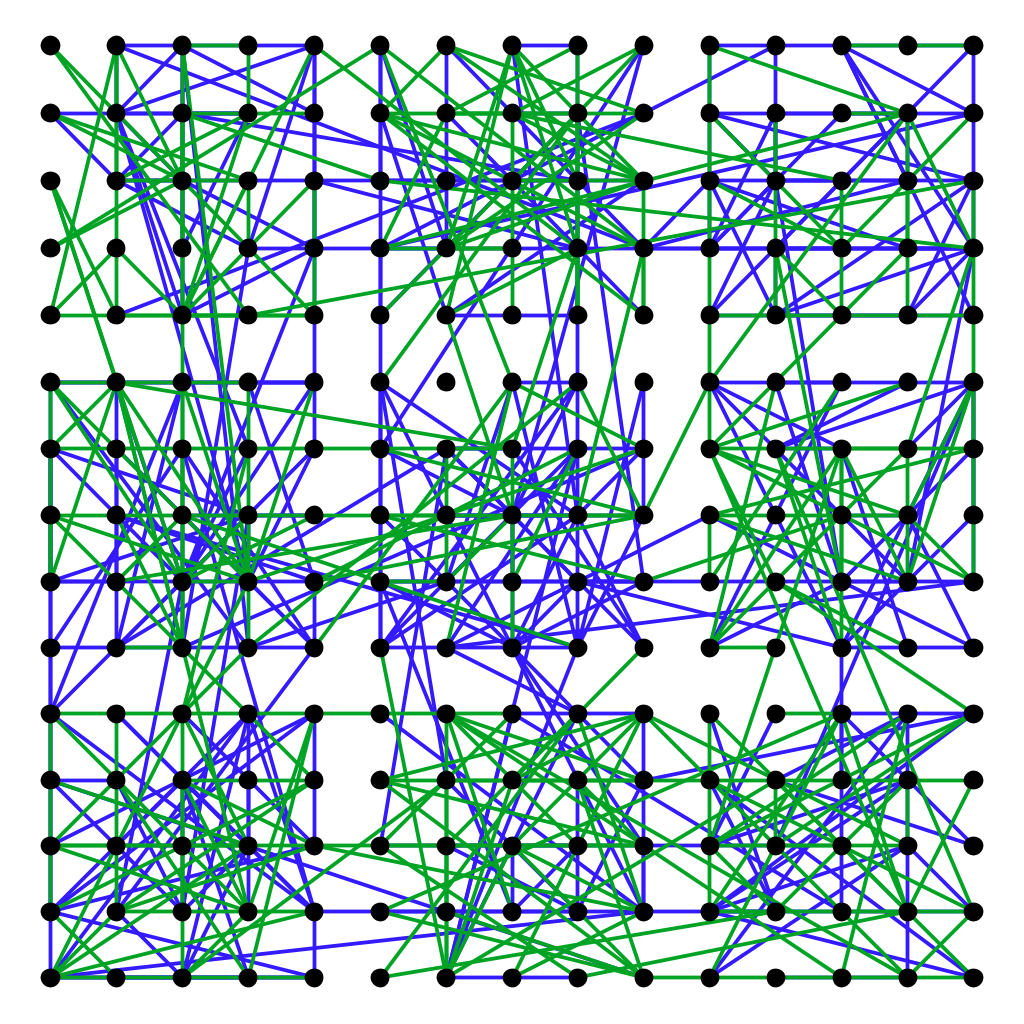}
	\caption{
		\textbf{A schematic representation of the model.} 
		The nodes are at the lattice sites of a two-dimensional square lattice of size $L\times L$ with $L = 15$.
		The system is constructed as $m\times m$ \er~(ER) networks.
		Here $m = 3$, where each ER is of size $\zeta\times \zeta$ with $\zeta = 5$.
		In each ER network the links are between pairs of sites chosen at random, and there are also links between nearby ER networks.
		The green and blue lines represent the links in the first and second layer of the multiplex.
		In our simulations we set periodic boundary conditions, 
		not shown for clarity.
		}
	\label{fig:Demonstration2}
\end{figure}

Here, we analyze and predict the resilience of spatial multiplex networks with modular structure, see Fig. \ref{fig:Demonstration2}, under localized attacks, by developing tools based on percolation theory. 
An example of a realistic system that motivates our model is the infrastructure networks in a country, where each layer describes different infrastructure.
The different infrastructures are dependent on each other, and in addition, each layer has high connections within the cities and a few long connections between nearby cities.
We focus on localized failures because of two main reasons. 
First, a localized damage is a realistic scenario (due to flood or earthquake), and second, in such systems, a finite number of local failures concentrated in the same area  might spread the damage throughout the system and cause significant damage and even to fully system collapse.


\textit{Model.}
Our model is generated as a multiplex system with spatial and community properties, see Fig. \ref{fig:Demonstration2}.
A multiplex network is a single network with at least two kinds of connectivity links. 
We assume here that the two types of links serve for two different functions, such as transportation and communication.
In fact, a multiplex network with two kinds of connectivity links (for instance) can be regarded as a special case of interdependent networks with two layers with the same number of nodes, and every node in one layer has only one interdependent link with a single node in the other layer. 
For a node to remain functional in the multiplex, after the percolation process \cite{staufferaharony,bunde1991fractals}, it must be connected to the giant component in both layers. 
This reflects the assumption that in order for a node in the system to function --- it requires both resources provided by the two layers.
	
Our multiplex model is composed here, for simplicity and without loss of generalization, of two layers in which the nodes are placed at sites of a square lattice of size $L\times L$.
The multiplex is constructed as $m\times m$ \er~(ER) multiplex networks (communities), each of which of size $\zeta\times\zeta$, that are tiled and connected to each other as a square lattice (see Fig. \ref{fig:Demonstration2}).
We assume, that the links within a community (intra-links) are connected at random, while the links connecting nodes in two distinct communities (inter-links) can only connect neighboring tiles.
Each node has a degree $\langle k_{inter} \rangle$ of inter-links and a degree $\langle k_{intra} \rangle$ of intra-links. The total degree is: $\ktot = \langle k_{inter} \rangle + \langle k_{intra} \rangle$.
In addition, the heterogeneity of the system is specified by the interconnectivity parameter $ \alpha = \langle k_{inter} \rangle / \ktot $.
It should be noted that the homogeneous case (without communities) has been previously studied both for single-layer \cite{grossvaknin-JPS,bonamassa-prl2019} and multi-layer \cite{danziger-epl2016,vaknin-njp2017} networks. In that model, all links have a characteristic length $\zeta$ with no distinction between inter and intra links, and therefore representing an homogeneous systems, without communities.
In contrast, the present new model can describe systems with a spatial structure of communities, where the heterogeneity of the system is controlled by the $\alpha$ parameter.
Thus, this model enables us to expand the previous model to a more general and realistic one for systems such as interconnected cities.


\begin{figure}
	\centering
	\includegraphics[width=\linewidth]{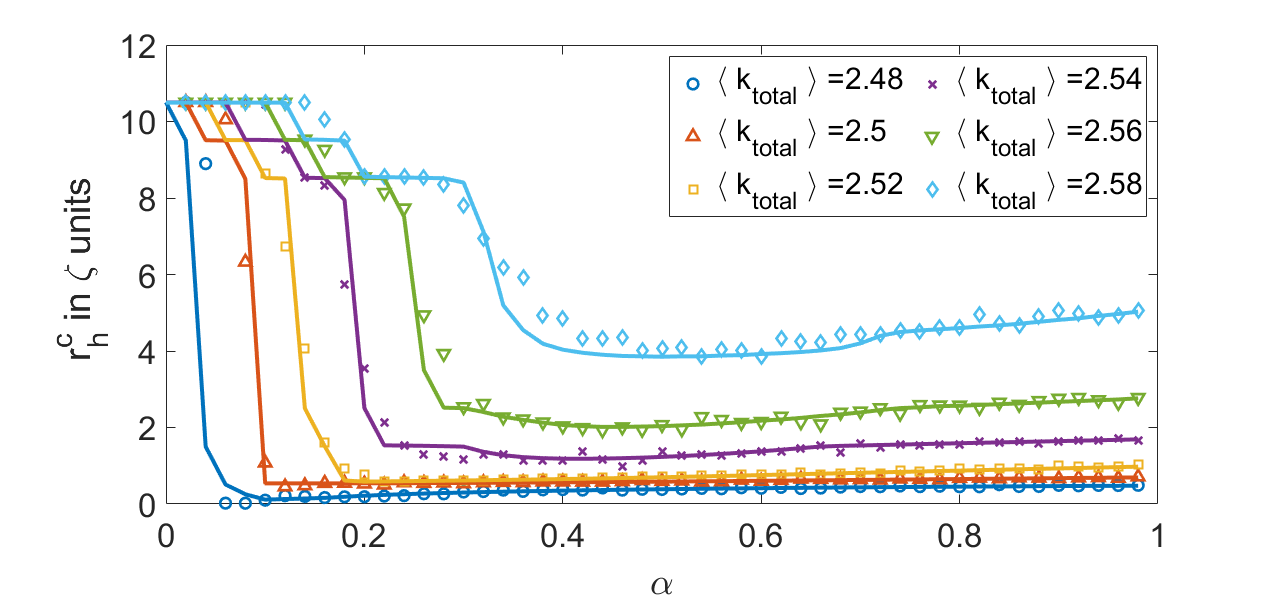}
	\caption{{\bf The critical attack size $r_h^c$ as a function of $\alpha$ for different $\ktot$ values}.
		For every  $\ktot$ the lines represent the theory of Eq. \eqref{eq:P_infty} and the symbols represent simulation results.
		As seen, the simulations show excellent agreement with the theory. 
		For a given $\ktot$, there is a metastable region that starts above a certain critical $\alpha_c$.
		In this region, $r_h^c$ is smaller then the system size and increases only slightly with $\alpha$.
		The figure suggests that for a certain $\ktot$ one needs $\alpha$ above a critical $\alpha_c$ to enable propagation of a localized damage. Below $\alpha_c$ one needs to remove order of $L^2$ nodes for the system to collapse.
		Above $\alpha_c$ a finite number of nodes (zero fraction) leads to system collapse.
		For the simulations we set $L = 2100$ and $\zeta=100$, with average over 5 runs for each data point.
	}
	\label{fig:plot_rc_alpha}
\end{figure}

\textit{Theory and Simulation Results.}
Here we develop a framework for understanding the cascading process for a general case of a multiplex network. The constrains are that each layer has a spatial structure of communities that are connected in a form of a lattice.
The cascading failures in multiplex networks is a process in which failures of some nodes lead to failures of other nodes and so on. When the cascading failures stops we are left with the mutual giant component ($MGC$) of the network. In the cascading process, at first we remove all nodes that are not in the giant component ($GC$) of the first layer. Then, from the set of the remaining nodes, we remove all the nodes that are not in the $GC$ of the second layer. We repeat this two steps until there are no nodes to remove, and we are left with the $MGC$.
The existence of a $MGC$ of volume $O(N)$ where $N$ is the number of nodes in the network, expresses the functionality of the system.

In order to obtain analytically the size of the $MGC$, we use a method similar to a finite element approach \cite{brenner-springscience2007} in which we introduce non-linear equations for each community and for each intercommunity link treating the entire system as a network of communities. We begin with deriving equations for the size of the $GC$ after a percolation process in a single layer and precisely defining the parameters of these equations.   
We assume that the number of links $k_{\nu,i,j}$ linking any node $\nu$ in community $i$ to nodes in community $j$ is statistically independent from number $k_{\nu,i,\ell}$,
linking node $\nu$ to any other community $\ell$. These numbers are randomly taken from a given degree distributions $P_{i,j}(k)$. We define 
the generating functions of these distributions:
\begin{equation}
G_{i,j}(x)=\sum_k P_{i,j}(k)x^k
\end{equation}
and the generating functions of the excess degree distribution \cite{newman-siam2003}:
\begin{equation}
H_{i,j}(x)=\sum_k\frac{ P_{i,j}(k) k}{\mkkij} x^k ,
\end{equation} 
where $\mkkij$ are the average degrees of distributions $ P_{i,j}(k)$.  
We define $\fij$ as the chance that a link passing from a node in community $i$ to a node in community $j$ does not lead to the $GC$. The link is an ``intra-link" if $i$ is equal to $j$ otherwise it is an ``inter-link". 
If we assume that $p_j$ is the fraction of nodes survived in community $j$ after an initial attack or as a result of cascading process, $\fij$ must satisfy recurrent equations:
\begin{equation}
\fij = (1-p_j) + p_j \cdot  H_{j,i}(f_{j,i}) \cdot \prod_{\ell \neq i} G_{j,\ell}(f_{j,\ell}),
\label{eq:lattice_fij}
\end{equation}
where the index $\ell$ goes over the set of neighboring communities of community $j$ including community $j$ itself. 
When $\ell=i$, one of the links leading from a node in community $j$ back to community $i$ is used by an incoming link, hence
we must use the generating function of the excess degree distribution.
Finally, the fraction of nodes in community $i$ which belong to the $GC$ is:
\begin{equation}
g_i=p_i\left[1- \prod_j G_{i,j}(\fij)\right],
\label{eq:lattice_gi}
\end{equation}
where the index $j$ goes over the set of neighboring communities of community $i$ including community $i$ itself.
If we introduce vectors $\vec{f}$ with components $\fij$, $\vec{p}$ with
components $p_i$ and $\vec{g}$ with components $g_i$,
then Eq. (\ref{eq:lattice_fij}) can be written in a symbolic vector form:
\begin{equation}
\vec{f}=\vec{\Phi}(\vec{f},\vec{p}).
\end{equation}
This equation can be solved using the iteration method starting with $\vec{f}=0$, and it will uniquely define the vector $\vec{f}(\vec{p})$ as function of vector $\vec{p}$.
Analogously, Eq. (\ref{eq:lattice_gi})
can be presented in a vector form:
\begin{equation}
\vec{g}=\vec{\Psi}(\vec{f},\vec{p}).
\label{e:Psi}
\end{equation}
For generality, we will assume that in Eq. (\ref{e:Psi}) the vectors $\vec{f}$ and $\vec{p}$, are two arbitrary vectors, independent of one another.
 
Now we will obtain equations for the $MGC$ of the multiplex. Suppose that the survival probability vector after initial
attack is $\vec{p}(0)$ with components $p_i(0)$, and the vector of
survival probabilities after stage $t$ of the cascade is $\vec{p}(t)$.
In principle, for the layers of the multiplex $A$ and $B$ the degree
distributions can be different, and hence the functions $\vec{\Phi}$ and
$\vec{\Psi}$ and the vectors $\vec{f}$ and $\vec{g}$ should be different. Therefore, we will distinguish them by adding indexes $A$ and
$B$.
Using the same logic as in Ref.~{\cite{buldyrev-nature2010}, the equations of the cascade of failures starting from $t=0$ will be:
	
\begin{equation}
	\label{eq:cascade}
	\begin{gathered}
		\vec{f}_A(2t)=\vec{\Phi}_A[\vec{f}_A(2t),\vec{p}(2t)]\\
		\vec{g}_A(2t)=\vec{\Psi}_A[\vec{f}_A(2t),\vec{p}(2t)]\\
		\vec{p}(2t+1)=\vec{\Psi}_A[\vec{f}_A(2t),\vec{p}(0)]\\
		\vec{f}_B(2t+1)=\vec{\Phi}_B[\vec{f}_B(2t+1),\vec{p}(2t+1)]\\
		\vec{g}_B(2t+1)=\vec{\Psi}_B[\vec{f}_B(2t+1),\vec{p}(2t+1)]\\
		\vec{p}(2t+2)=\vec{\Psi}_B[\vec{f}_B(2t+1),\vec{p}(0)],
	\end{gathered}
\end{equation}	
where $\vec{g}_A(t)$ and $\vec{g}_B(t)$ are the fraction of nodes of each community in the giant component at stage $t$ and $\vec{p}(t)$ is the effective fraction of survived nodes representing stage $t$ of the cascade of failures as a percolation process after a random attack.  As $t\to\infty$ the vectors $\vec{g}_A(t)$ and $\vec{g}_B(t)$ will converge to the mutual giant component $\vec{P}_\infty$.

If all distributions $P_{i,j}(k)$ are Poisson distributions as in ER graphs, then $G_{i,j}(x)=H_{i,j}(x)=\exp[\mkkij(x-1)]$ and all probabilities $\fij$ for the same community $j$ but different $i$ satisfy the same equation and hence they must be equal and we define $f_j\equiv f_{i,j} $.
Thus, Eq. (\ref{eq:lattice_fij}) and Eq. (\ref{eq:lattice_gi}) are significantly simplified and become respectively:
\begin{equation}
f_j = (1-p_j) + p_j \cdot  e^{\sum_{i} \mkkij \cdot (f_i -1)}
\label{eq:lattice_fj}
\end{equation}
and
\begin{equation}
g_j = p_j(1-e^{\sum_{i} \mkkij \cdot (f_i -1)}). 
\label{eq:lattice_gj}
\end{equation}
From this follows that $g_j=(1-f_j)=1-\exp(-\sum_i\mkkij g_i)$.  
When the two layers of multiplex A and B are with average degrees $\mkkij_A$, and $\mkkij_B$ respectively, Eqs. (\ref{eq:cascade}) give:
\begin{equation}
\begin{aligned}
{P_\infty}_j=&~p_{0_j}\cdot[1-\exp(-\sum_i\mkkij_A {P_\infty}_i)]\cdot\\ &[1-\exp(-\sum_i\mkkij_B {P_\infty}_i)],  
\label{eq:P_infty}
\end{aligned}
\end{equation}
which is a generalization of the mutual giant component formula for a multiplex of homogeneous graphs given in Eq. (40) of Ref. \cite{gao-pre2012}.  We find the numerical solution of the system (\ref{eq:P_infty}) by iterations starting from the initial values ${P_\infty}_j=1$.



\begin{figure}
	\centering
	\subfloat[]{\includegraphics[width=0.5\linewidth]{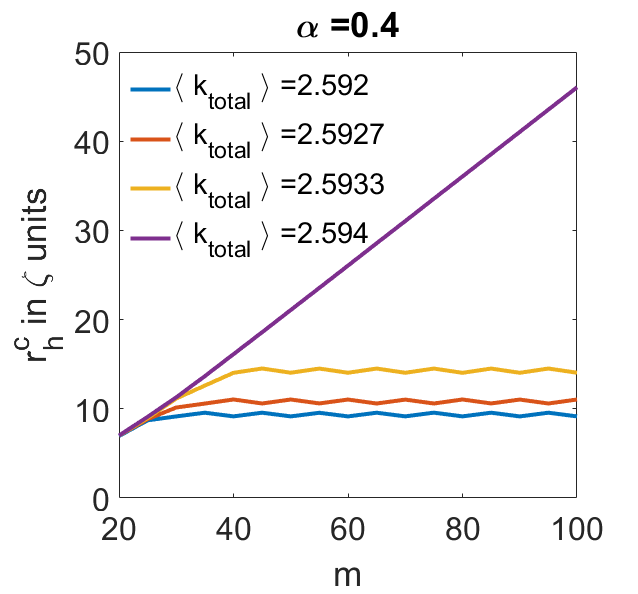}}
	\subfloat[]{\includegraphics[width=0.5\linewidth]{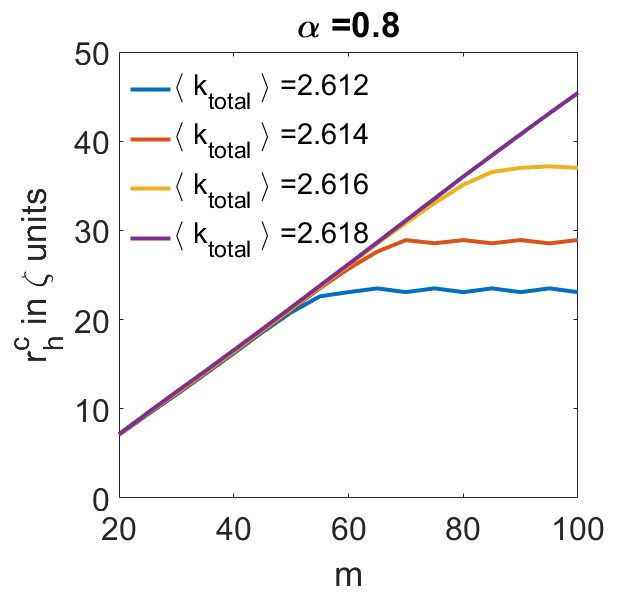}}\\
	\subfloat[]{\includegraphics[width=\linewidth]{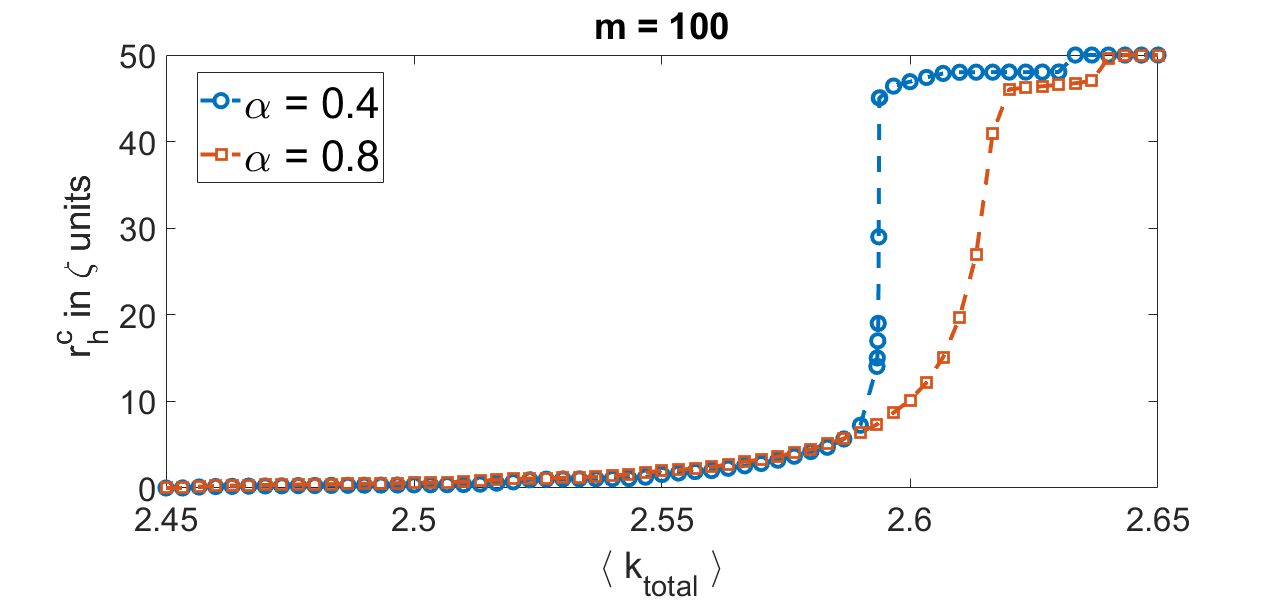}}
	\caption{\textbf{Analytical results --- the dependence of the critical attack size $r_h^c$ on the number of communities.}
		In (a) and (b) we show three behaviors of $r_h^c$  for $\alpha = 0.4$ and $\alpha = 0.8$. The number of communities is $m \times m$,  therefore the dependence on $m$ expresses the dependence of $r_h^c$ on the number of communities. For relatively small $\ktot$, $r_h^c$ does not depend on the number of communities. For intermediate values of $\ktot$,  $r_h^c$ initially grows with $m$ and then from a certain $m$, reaches a stable value. This stable value oscillates between two values of $r_h^c$ which correspond to even and odd values of $m$. For large $\ktot$ values,  $r_h^c$ grows linearly with $m$ and is approximately $0.5m$.
		In (c) we show (for $m = 100$) that $r_h^c$ increases with $\ktot$, when for $\alpha = 0.4$ the transition is sharper than for $\alpha = 0.8$. This result  brightens why the jump in (a) larger than in (b).
	}
	\label{fig:rcm_alpha0408_secondoption}
\end{figure}

\begin{figure}
	\centering
	\subfloat[]{\includegraphics[width=\linewidth]{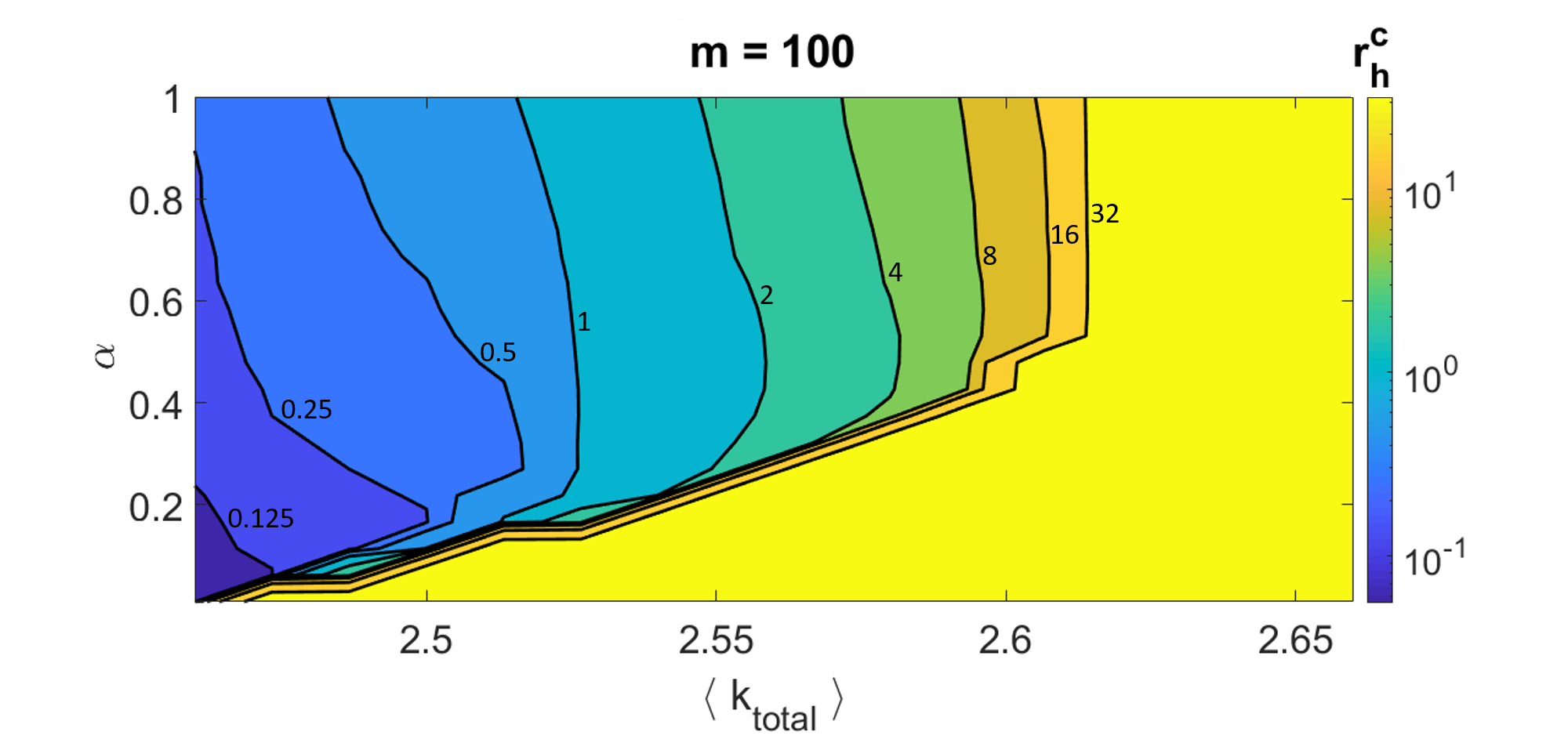}}\hfil
	\subfloat[]{\includegraphics[width=\linewidth]{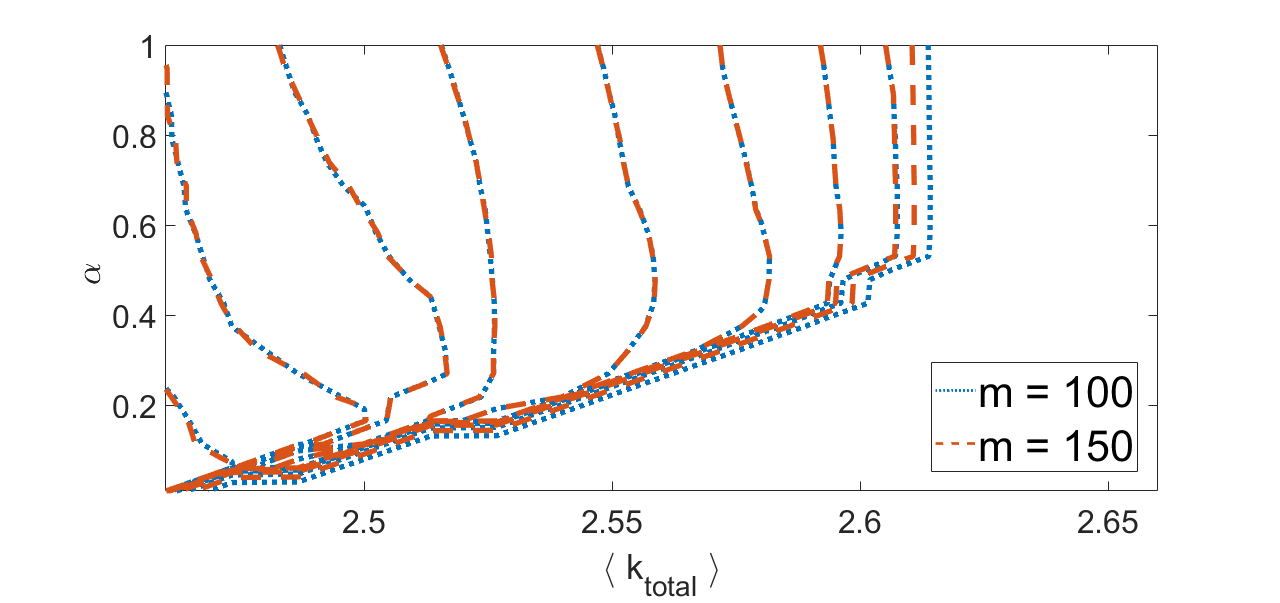}}
	\caption{\textbf{Analytical results --- Phase diagram of the critical attack size $r_h^c$.}
		Dependence of $r_h^c$ on the average degree $\ktot$ and the interconnectivity parameter $\alpha$.
		For the phase diagrams in this figure, we sample with equal intervals $16$ values for $\ktot$ and $20$ values for $\alpha$.
		(a) Contour for the phase diagram with $m = 100$. The color bar in the right represents the size of $r_h^c$ in $\zeta$ units (in log scale).
		(b) Here we show the same contour lines as in (a) for $r_h^c$ values for $m = 100$ and $m = 150$. We see that both $m$ values give identical results except for near to the border where $r_h^c \sim 0.5 L$ (in the last contour line).
		For lower $m$ values, the misfit of the contour lines starts in lower values of $r_h^c$.
	}
	\label{fig:rck}
\end{figure}


\begin{figure}
	\centering
	\subfloat[]{\includegraphics[width=\linewidth]{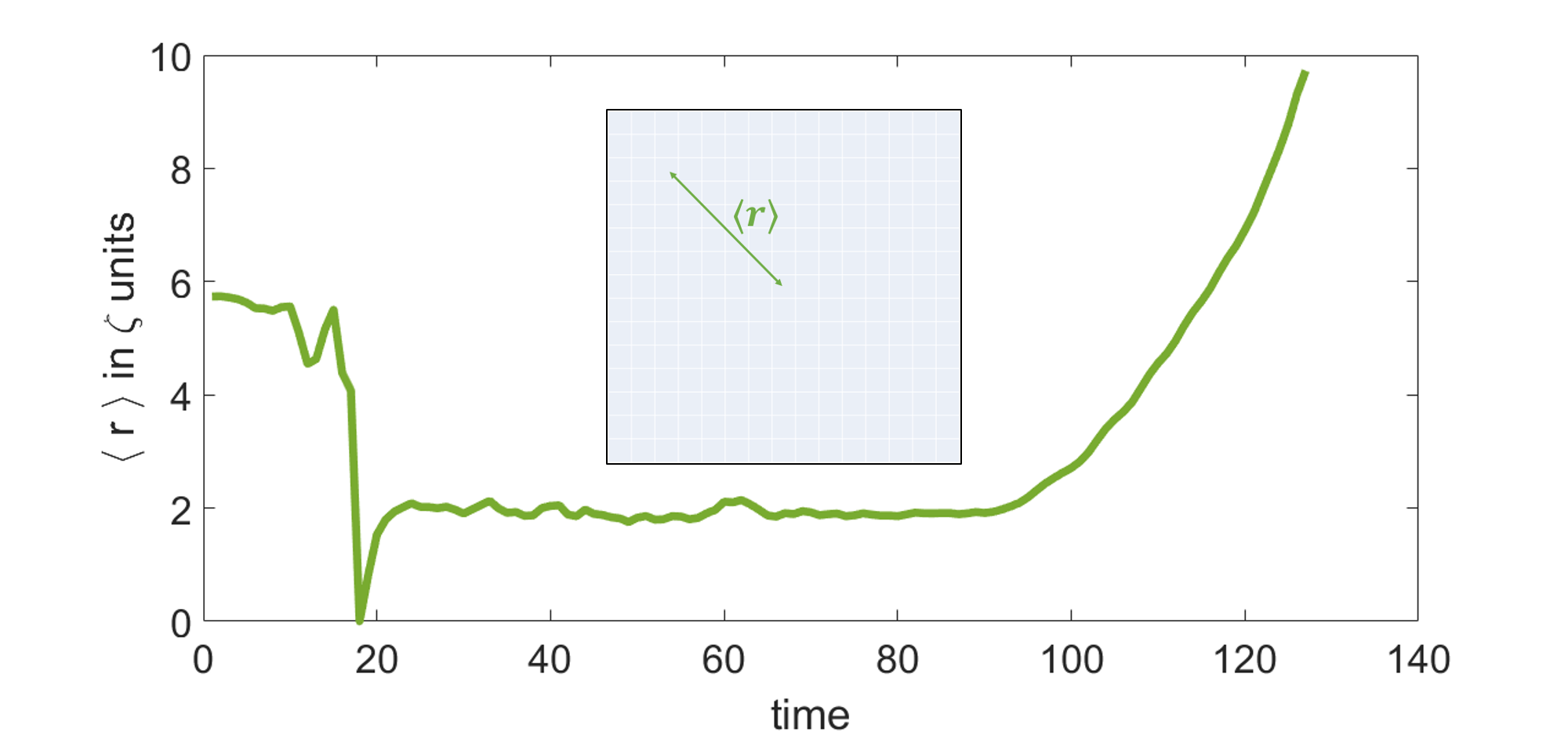}}\hfil
	\subfloat[]{\includegraphics[width=\linewidth]{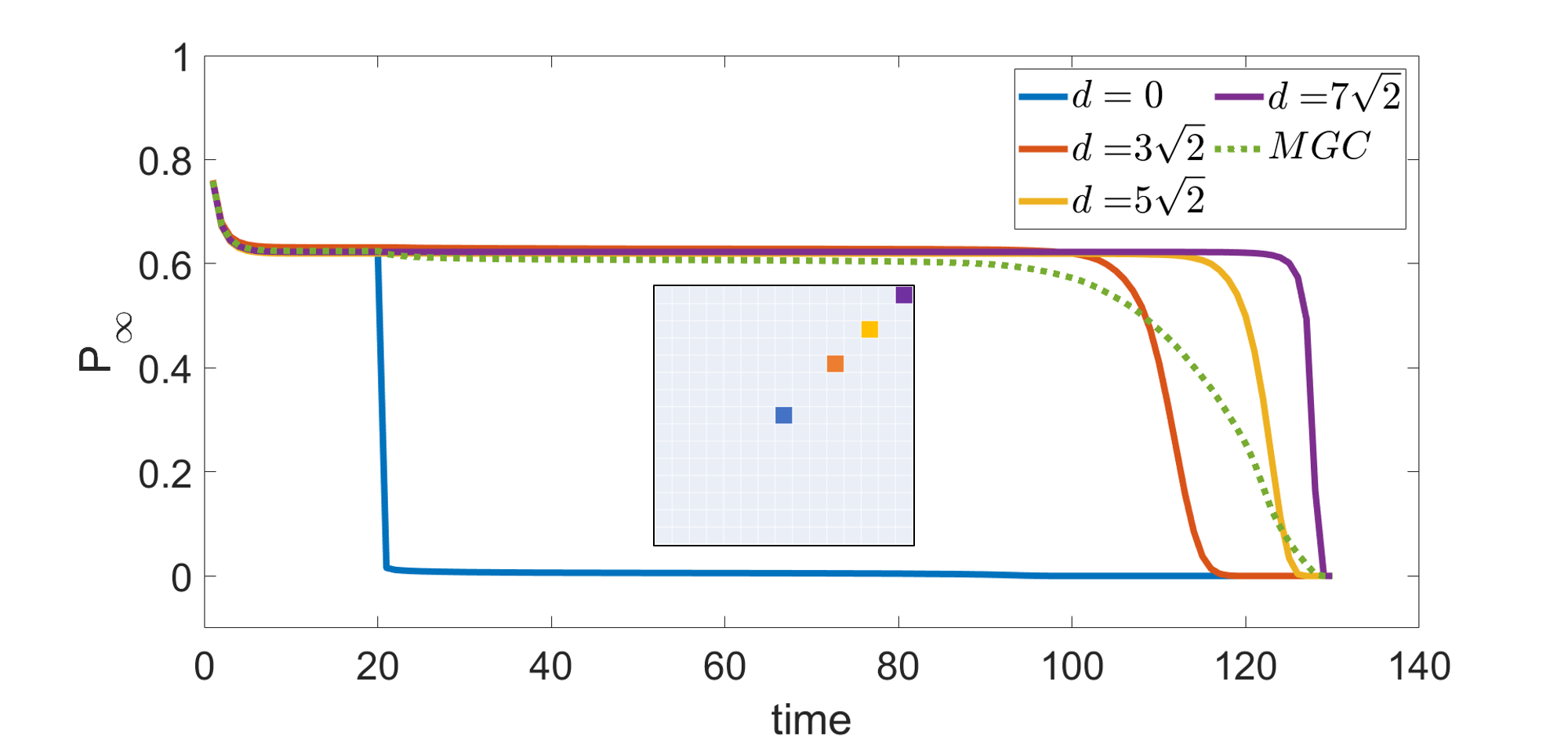}}
	\caption{\textbf{The cascading failures near the critical point.}
		Propagation of a local damage with radius slightly above the critical size $r_h^c$.
		(a) The average distance from the center, $\langle r \rangle$, of the nodes that fail at every iteration.
		The inset figure is an illustration of the network.
		(b) The continuous lines represent the size of ${P_\infty}_j$, for $4$ communities having different distances $d$ from the center (where the critical hole was removed), as a function of time.
		The colors of the lines correspond to the colors of the painted communities in the inset figure.
		The dotted line shows the $MGC$ size over the whole multiplex ($\sum_{j}{P_\infty}_j$).
		For the simulations we set $L = 4500$, $\zeta=300$, $\ktot = 2.5$ and  $\alpha = 0.4$. The critical size $r_h^c$ for this simulation, which obtained through a binary search, is $r_h^c = 0.57$ in $\zeta$ units.}
	\label{fig:age}
\end{figure}

We next analyze the robustness of our community multiplex model with respect to localized attacks or failures. 
To this end, we consider the case where all nodes within a radius $r_h$ (radius-hole), from the center of the multiplex, are removed from the network. 
When $m$ is an even number, then the center of the multiplex is at the corner of 4 neighboring ER communities, and else it is in the center of one ER. Note that $r_h$ translates into the value of $p_i(0)$ by counting the fraction of lattice sites outside the hole of radius $r_h$ in the damaged communities. For example, if the center of the hole is in the center of the community $i$ and $r_h<1/2$, $p_i(0) \approx 1-\pi r_h^2$, alternatively, when the center is at the corner of the four communities $j$, $p_j(0)\approx 1-\pi r_h^2/4$ in each of the four damaged communities.
    
We find for networks with different system parameters, by simulations and theory, what is the critical radius $r_h^c$ needed to cause a system collapse. We find the accurate value of $r_h^c$ through a binary search, where increasing or decreasing of the radius attack is determined by the $MGC$ size at the end of the cascading. At a given radius attack $r_h$, if the $MGC$ size (after the cascading) remains in order of the system size then we increase $r_h$, and otherwise -- we decrease it.
We define a threshold condition for the $MGC$ size, below which we assume that the $MGC$ is 0.
For the numerical calculations of the theory we set the threshold to be $10^{-12}$, and for the simulations (after some tests) a fraction of $0.1$ of the system size seems to provide a good threshold condition.
Furthermore, for the numerical calculations we divide each community into $\zeta^2$ of points. 
Thus, we can calculate numerically with good approximation the fraction of nodes that fail in each community for an attack $r_h$. In addition, since we study the case of a symmetrical two-dimensional square lattice, for the theoretical calculations (using Eq. \eqref{eq:P_infty}) we set $\mkkij_A$ and $\mkkij_B$ to be $\kinter /4$ for $i\neq j$.

We find that for a network with structure parameters within a certain parameter range of $L, \zeta$ and $\langle k_{total} \rangle$, there are two regimes that are divided by a critical $\alpha_c$, see Fig. \ref{fig:plot_rc_alpha}.
Hence, a different ratio $\alpha$ between $\langle k_{inter} \rangle$ and $\ktot$ --- for a fixed $\ktot$ --- can completely change the system's resilience to localized attacks.
For $\alpha > \alpha_c$, we have a metastable regime, where a finite size localized attack larger than $r_h^c$ causes cascading failures, leading to system collapse. The critical radius $r_h^c$ in this regime, for a given $\langle k_{total} \rangle$, depends weakly on the interconnectivity parameter $\alpha$. Note that this metastable regime located in the narrow interval of $\ktot$ above $k_c \approx 2.4554$, where $k_c$ is the critical average degree (independent of $\alpha$) below which the homogeneous ER multiplex collapses without any initial damage \cite{buldyrev-nature2010}. In order to interpret our model as a model of a real infrastructure, we must assume that the infrastructure is designed to be as economical as possible: i.e. $\langle k_{total}\rangle$ is minimized in a such a way that the multiplex retains the mutual giant component.  Moreover, we assume that the initial state of the system before the localized attack is the MGC of the multiplex with given initial $\langle k_{total}\rangle$. Since in the mutual percolation, the final state of the system does not depend on the order in which the damage was made, the
final result is the same as if the localized attack was produced simultaneously with the creation of the multiplex as given by Eqs. (\ref{eq:cascade}). Remarkably, in this metastable regime 
networks with the same $\langle k_{total}\rangle$ but larger interconnectivity ratio $\alpha$ are more vulnerable to localized attacks than networks with small $\alpha$ where the communities are not well connected, but more self-sufficient.
Moreover, in this metastable regime, $r_h^c$ is independent of the number of communities (Fig. \ref{fig:rcm_alpha0408_secondoption}).
In marked contrast, for $\alpha < \alpha_c$, the critical attack $r_h^c$ is $\sim0.5m$ in $\zeta$ units. Namely, the system remains functioning for any attack that is less than removing the entire system.
In addition, we obtain numerically based on Eq. \eqref{eq:P_infty} a phase diagram of $r_h^c$, for a large system with $m = 100$ in Fig. \ref{fig:rck} (a). 
The phase diagram is the same for different $m$ values, see for instance Fig. \ref{fig:rck} (b), except for $r_h^c$ that are in the order of the system size.

As we noted above, the initial state of the system must be understood as the MGC obtained for given $\langle k_{total}\rangle$ before the localized attack. Therefore, to understand how the damage produced by the localized attack spreads with time, we first produce the cascade of failures with $p_i(0)=1$. After this cascade stops we produce the localized attack of a given radius $r_h^c$. For example, for the simulations in Fig. \ref{fig:age} we perform the attack on step $time = 19$. After the attack, there is a long latent period during which only a few nodes fail at every time step, and they are located mostly in the vicinity of the attack area. Then, the damage quickly spreads until it reaches the edges of the system. The spreading process explains why the attack size does not depend on the system size.

\textit{Discussion.}
Typically, real-world engineered systems are designed to be as efficient and cost effective as possible. Therefore, such systems are slightly above the critical state, and hence, these systems are very vulnerable to various local failures.
Here we have investigated the stability of realistic interdependent networks, consisting of interconnected communities embedded in space, against local failures. We develop a theory for calculating the magnitude of the critical damage needed to destroy the entire system for different parameters of connectivity and spatiality. 
Our approach is similar to the finite element method which is applied here to the network of communities, where each community is treated as an element, participating in a system of equations.
We find that for the same $\langle k_{total} \rangle$ (and, hence, the same cost) the networks with low interconnectivity $\alpha$ are more robust against localized attacks than the system in which the communities are well connected. If $\alpha$ is large, the damage produced by the localized attack spreads over the entire system. For small $\alpha$, the damage does not spread. Thus, the interlinks connecting neighboring communities could serve as vehicles of damage propagation rather than for stabilizing the system. This finding explains why islanding, the strategy employed the electrical engineers by dividing the system into almost isolated self-sustained islands is an efficient strategy against cascading failures in the power grid. 
In addition, we study the dynamical process of cascading and find a long latent period during which the number of failed nodes is very small and they are localized close to the initial attack. 
During this period, a relatively small intervention by reinforcing a few nodes can stop the propagation of the cascade of failures.  After the latent period is over, the damage quickly spreads over the entire system and there is no economic way to stop it.

\textit{Acknowledgement.}
We thank the Italian Ministry of Foreign Affairs and International Cooperation jointly with the Israeli Ministry of Science, Technology, and Space (MOST); the Israel Science Foundation, ONR, the Japan Science Foundation with MOST, BSF-NSF, ARO, the BIU Center for Research in Applied Cryptography and Cyber Security, and DTRA (Grant no. HDTRA-1-14-1-0017) for financial support.
D. V. thanks the PBC of the Council for Higher Education of Israel for the Fellowship Grant.


\FloatBarrier
\bibliographystyle{naturemag_4etal}
\bibliography{mybib}
\end{document}